\documentclass[twocolumn,showpacs,preprintnumbers,amsmath,amssymb]{revtex4}
\usepackage{graphicx}% Include figure files
\usepackage{dcolumn}% Align table columns on decimal point
\usepackage{bm}% bold math

\begin{document}

%\preprint{APS/123-QED}

\title{Maximal violation of Mermin's inequalities}
% Force line breaks with \\
\author{Zeqian Chen}
\email{zqchen@wipm.ac.cn}
\affiliation{%
Wuhan Institute of Physics and Mathematics, Chinese Academy of
Sciences, P.O.Box 71010, 30 West District, Xiao-Hong Mountain,
Wuhan 430071, China}%

\date{\today}% It is always \today, today,

\begin{abstract}
In this paper, it is proved that the maximal violation of Mermin's
inequalities of $n$ qubits occurs only for GHZ's states and the
states obtained from them by local unitary transformations. The
key point of our argument involved here is by using the certain
algebraic properties that Pauli's matrices satisfy, which is based
on the determination of local spin observables of the associated
Bell-Mermin operators.
\end{abstract}

\maketitle

The Bell inequality [1] was originally designed to rule out
various kinds of local hidden variable theories. Precisely, the
Bell inequality indicates that certain statistical correlations
predicted by quantum mechanics for measurements on two-qubit
ensembles cannot be understood within a realistic picture based on
Einstein, Podolsky, and Rosen's (EPR's) notion of local realism
[2]. However, Bell-type inequalities also provide tests to
distinguish entangled from nonentangled quantum states. For
example, Gisin's theorem [3] asserts that all entangled two-qubit
states violate the Clauser-Horne-Shimony-Holt (CHSH) inequality
[4] for some choice of spin observables, that is, the violation of
the CHSH inequality characterizes entangled states of two qubits.

As is well known, maximally entangled states, such as Bell states
and GHZ states [5], have become a key concept in the nowadays
quantum mechanics. On the other hand, from a practical point of
view maximally entangled states have found numerous applications
in quantum information [6]. A natural question is then how to
characterize maximally entangled states. There are extensive
earlier works on maximally entangled states [7], however, this
problem is far from being completely understood today. It is
argued that the maximal violation of the Bell-type inequalities
can be used to characterize maximally entangled states [8].
Therefore, for characterizing maximally entangled states, it is
suitable to study the states that maximally violate the Bell-type
inequalities. Recently, the author [9] has completely resolved the
two-qubit case and shown that the Bell states and the states
obtained from them by local unitary transformations are the unique
states that violate maximally the CHSH inequality.

It is natural to consider the $n$-qubit case in terms of Bell-type
inequalities. This is not precise, because there are infinitely
many versions of the Bell-type inequalities [10]. It is well known
that Mermin [11] derived an $n$-particle Bell-type inequality,
which provides the first spectacular demonstration of the fact
that there is no limit to the amount by which quantum-mechanical
correlations can exceed the limits imposed by the premises of
EPR's local realism. In this paper, we will show that the maximal
violation of Mermin's inequalities characterizes the GHZ states of
$n$-qubit as similar to the two-qubit case in terms of the CHSH
inequality, that is, the GHZ states and the states obtained from
them by local unitary transformations are the unique states that
violate maximally Mermin's inequalities.

The techniques involved here are introduced by the author [9] and
based on the determination of local spin observables of the
associated Bell operator. We show that a Bell-Mermin operator
presents a maximal violation on a state if and only if the
associated local spin observables satisfy the certain algebraic
identities that Pauli's matrices satisfy. Consequently, we can
find those states that show maximal violation, which are just the
states obtained from the GHZ states by local unitary
transformations. Our method involved here is simpler (and more
powerful) than one used in [12], which is based on the
determination of the eigenvectors and eigenvalues of the
associated Bell operator. We have illustrated the three-qubit case
in [9] and shown that all states that maximally violate the
Bell-Klyshko inequality [13] are exactly GHZ states and the states
obtained from them by local unitary transformations, which was
conjectured by Gisin and Bechmann-Pasquinucci [8].

Let us give a brief review of Mermin's inequality of $n$ qubits.
Let $A_j,A'_j$ denote two spin observables on the $j$th qubit,
$j=1,...,n.$ Define the Bell-Mermin operator as
following\begin{equation}{\cal M}_n = \frac{1}{2i} \left (
\bigotimes^n_{j=1} (A_j + i A'_j ) - \bigotimes^n_{j=1} (A_j - i
A'_j ) \right ).\end{equation}It is easy to check
that\begin{equation}\begin{array}{lcl}{\cal M}_n &=&A'_1A_2 \cdots
A_n+ A_1A'_2A_3 \cdots A_n + \cdots\\&~&-A'_1A'_2A'_3A_4 \cdots
A_n - A'_1A'_2A_3A'_4 \cdots A_n - \cdots\\&~&+A'_1 \cdots A'_5A_6
\cdots A_n + \cdots\\&~&\cdots,\end{array}\end{equation}where each
line of Eq.(2) contains all distinct permutations of the
subscripts that give distinct products. Since the total number of
terms in Eq.(2) is $\sum_{j~odd} (^n_j)=2^{n-1},$ it concludes
that\begin{equation}\| {\cal M}_n \| \leq
2^{n-1}.\end{equation}However, assuming ``local realism" [2] one
concludes by using Mermin's elegant technique [11] that for $n$
even,\begin{equation}\left | \langle {\cal M}_n \rangle \right |
\leq 2^{n/2},\end{equation} and\begin{equation}\left | \langle
{\cal M}_n \rangle \right | \leq 2^{(n-1)/2},\end{equation}when
$n$ is odd, respectively. The inequalities appearing in Eqs.(4)
and (5) are said to be Mermin's inequalities. The original
Mermin's inequality is the case that $A_j = \sigma^j_x$ and $A'_j
= \sigma^j_y.$ Clearly, for $n=2$ there is no violation of
Mermin's inequalities for the prediction that quantum mechanics
makes. However, for $n \geq 3$ Mermin [11] showed that the
prediction $2^{n-1}$ that quantum mechanics makes for the GHZ
state\begin{equation}|GHZ \rangle = \frac{1}{\sqrt{2}} \left ( |
0\cdots 0 \rangle + i| 1\cdots 1 \rangle
\right),\end{equation}maximally violates the original Mermin's
inequality by an exponentially large factor of $2^{(n-2)/2}$ for
$n$ even or $2^{(n-1)/2}$ for $n$ odd. This is the first
spectacular demonstration of the fact that there is no limit to
the amount by which the quantum-mechanical correlations can exceed
the limits imposed by Bell-type inequalities.

In the sequel, we always assume $n \geq 3.$ The main result we
shall prove is that

{\it Theorem: A state $| \varphi \rangle$ of $n$ qubits maximally
violates Eq.$(4)$ for $n$ even or Eq.$(5)$ for $n$ odd, that
is,\begin{equation}\langle \varphi | {\cal M}_n | \varphi \rangle
= 2^{n-1},\end{equation}if and only if it can be obtained by a
local unitary transformation of the GHZ state $|GHZ \rangle,$
i.e.,\begin{equation}| \varphi \rangle = U_1 \otimes \cdots
\otimes U_n |GHZ \rangle\end{equation}for some $n$ unitary
operators $U_1,...,U_n$ on ${\bf C}^2.$}

The sufficiency of {\it Theorem} is clear. Indeed, as shown by
Mermin [11], the GHZ state $|GHZ \rangle$ satisfies Eq.(7) with
$A_j = \sigma^j_x$ and $A'_j= \sigma^j_y.$ For generic states $|
\varphi \rangle$ of the form Eq.(8), they maximally violate
Mermin's inequality of $n$ qubits with $A_j = U_j \sigma^j_x
U^*_j$ and $A'_j = U_j \sigma^j_y U^*_j,$ where $U^*_j$ is the
adjoint operator of $U_j.$

It remains to prove the necessity. Since\begin{equation}{\cal M}_3
= A'_1 A_2 A_3 + A_1 A'_2 A_3 + A_1 A_2 A'_3 - A'_1 A'_2
A'_3\end{equation}coincides with the Bell-Hardy-Klyshko operator
[13], the three-qubit case has been proved by the author [9]. The
key point of our argument involved in [9] is by using the certain
algebraic properties that Pauli's matrices satisfy, which is based
on the determination of local spin observables of the associated
Bell operator. As follows, by using this method we show that every
state $| \varphi \rangle$ satisfying Eq.(7) can be obtained by a
local unitary transformation of the GHZ state $|GHZ \rangle.$

Let us fix some notation. For $A^{(\prime)}_j=
\vec{a}^{(\prime)}_j \cdot \vec{\sigma}_j$ $(1\leq j \leq n),$ we
write$$(A_j,A'_j) = (\vec{a}_j, \vec{a}'_j ), A_j \times A'_j = (
\vec{a}_j \times \vec{a}'_j) \cdot \vec{\sigma}_j.$$Here
$\vec{\sigma}_j$ is the Pauli matrices for the $j$th qubit; the
norms of real vectors $\vec{a}^{(\prime)}_j$ in ${\bf R}^3$ are
equal to $1.$ Clearly,\begin{equation}A_j A'_j = (A_j,A'_j) + i
A_j \times A'_j,\end{equation}\begin{equation} A'_jA_j = (A_j,
A'_j) - i A_j \times A'_j,\end{equation}and\begin{equation}\| A_j
\times A'_j \|^2=1- (A_j, A'_j)^2.\end{equation}We always write
\begin{equation}A''_j = A_j \times A'_j.\end{equation}
By Eqs.(10), (11) and (13), one has that\begin{equation}\left
[A_j,A'_j \right ] = 2i A''_j, \left \{A_j, A'_j \right \} = 2
(A_j, A'_j)\end{equation}where $\left [A_j,A'_j \right ]$ and
$\left \{A_j, A'_j \right \}$ are the commutator and
anticommutator of the spin observables $A_j$ and $A'_j,$
respectively.

As follows, we write $A''_{j_1}A''_{j_2},$ etc., as shorthand for
$$I\otimes \cdots I\otimes A''_{j_1} \otimes I \otimes \cdots I
\otimes A''_{j_2}\otimes I \cdots \otimes I,$$where $I$ is the
identity on a qubit. The proof of {\it Theorem} for necessity
consists of the following three steps:

(i)  At first, we show that if $\| {\cal M}^2_n \| = 2^{2(n-1)}$
for $n \geq 3,$ then\begin{equation}(A_j,A'_j) =0\end{equation}for
all $j=1,...,n.$

We need to distinguish between the cases of $n$ odd and $n$ even.
For $n$ odd ($n \geq 3$), the square of the Bell-Mermin's operator
Eq.(2) is that\begin{widetext}$${\cal M}^2_n =
\sum^{(n-1)/2}_{k=0} (-1)^k 2^{n-2k-1}\sum_{1 \leq j_1 < j_2
<\cdots < j_{2k}\leq n} \left [A_{j_1},A'_{j_1} \right ]\left
[A_{j_2},A'_{j_2} \right ]\cdots \left [A_{j_{2k}},A'_{j_{2k}}
\right ],$$where $\{ j_1, j_2,\ldots, j_{n-1} \}$ is a set of
$n-1$ indices each of which runs from $1$ to $n.$ Hence, by
Eq.(14) one has that\begin{equation}{\cal M}^2_n = 2^{n-1}
\sum^{(n-1)/2}_{k=0} \left ( \sum_{1 \leq j_1 < j_2 < \cdots <
j_{2k} \leq n} A''_{j_1} A''_{j_2} \cdots A''_{j_{2k}} \right
).\end{equation}Analogously, for $n$ even ($n \geq 4$) one has
that\begin{equation}{\cal M}^2_n = 2^{n-1} \sum^{n/2}_{k=0} \left
( \sum_{1 \leq j_1 < j_2 < \cdots < j_{2k} \leq n} A''_{j_1}
A''_{j_2} \cdots A''_{j_{2k}} \right )- (-1)^{n/2} 2^{n-1}
(A_1,A'_1)(A_2,A'_2)\cdots (A_n,A'_n).\end{equation}Since
$\sum^{(n-1)/2}_{k=0} (^n_{2k})=2^{n-1},$ it immediately concludes
from Eqs.(12) and (16) that if $\| {\cal M}^2_n \| = 2^{2(n-1)}$
for $n \geq 3$ odd, then Eq.(15) holds true for all $j=1,...,n.$
Also, for $n$ even ($n \geq 4$), by Eqs.(12) and (17) one has
that$$\|{\cal M}^2_n \| \leq 2^{n-1}\left (1 - (-1)^{n/2}x_1\cdots
x_n + \sum^{n/2}_{k=1} \sum_{1 \leq j_1 < j_2 < \cdots < j_{2k}
\leq n} \sqrt{(1-x^2_{j_1})\cdots (1-x^2_{j_{2k}})}\right
),$$\end{widetext}where $x_j =(A_j,A'_j)$ for $j=1,...,n.$ It is
easy to check that when $-1 \leq x_j \leq 1$ for $j=1,...,n,$ the
function in the right hand of the above inequality attains the
maximal value $2^{2(n-1)}$ only at $x_1=\cdots =x_n=0.$ This
concludes Eq.(15).

In particular, if Eq.(15) holds true, it concludes that
\begin{equation}A_j A'_j=-A'_jA_j=iA''_j,\end{equation}
\begin{equation}A'_jA''_j=-A''_jA'_j=iA_j,\end{equation}
\begin{equation}A''_jA_j =-A_jA''_j=iA'_j,\end{equation}
\begin{equation}A^2_j=(A'_j)^2=(A''_j)^2 = 1,\end{equation}
i.e., $\{A_j,A'_j,A''_j\}$ satisfy the algebraic identities that
Pauli's matrices satisfy [14]. Therefore, by choosing
$A''_j$-representation $\{|0 \rangle_j,|1 \rangle_j \},$
i.e.,\begin{equation}A''_j |0 \rangle_j = |0 \rangle_j, A''_j |1
\rangle_j = -|1 \rangle_j,\end{equation}we have
that\begin{equation}A_j |0 \rangle_j =e^{-i \alpha_j} |1
\rangle_j, A_j |1 \rangle_j = e^{i \alpha_j} |0
\rangle_j,\end{equation}
\begin{equation}A'_j |0
\rangle_j = ie^{-i \alpha_j} |1 \rangle_j, A'_j |1 \rangle_j =-i
e^{i \alpha_j} |0 \rangle_j,\end{equation}for some $0\leq \alpha_j
\leq 2 \pi.$ We write $|0\cdot \cdot \cdot 0 \rangle_n,$ etc., as
shorthand for $|0 \rangle_1\otimes \cdot \cdot \cdot \otimes |0
\rangle_n.$ Then, $\{|\epsilon_1 \cdot \cdot \cdot \epsilon_n
\rangle_n : \epsilon_1,..., \epsilon_n = 0,1 \}$ is a orthogonal
basis of the n-qubit system.

(ii) Secondly, we prove that for $n \geq 3,$ a state $| \varphi
\rangle$ of $n$ qubits satisfying\begin{equation} {\cal M}^2_n |
\varphi \rangle = 2^{2(n-1)}| \varphi \rangle\end{equation}must be
of the form\begin{equation}| \varphi \rangle = a| 0\cdot \cdot
\cdot 0 \rangle_n + b| 1\cdot \cdot \cdot 1
\rangle_n,\end{equation}where $|a|^2 + |b|^2 =1.$

In this case, $\| {\cal M}^2_n \| = 2^{2(n-1)}$ and hence Eq.(15)
holds. By Step (i), we can uniquely write
$$| \varphi \rangle = \sum_{\epsilon_1,..., \epsilon_n = 0,1}
\lambda_{\epsilon_1 \cdots \epsilon_n }|\epsilon_1 \cdots
\epsilon_n \rangle_n,$$ where $\sum |\lambda_{\epsilon_1 \cdots
\epsilon_n}|^2 = 1.$ In particular, by Eq.(16) and (17) we have
that\begin{equation}A''_{j_1} A''_{j_2} \cdots A''_{j_{2k}} |
\varphi \rangle = | \varphi \rangle\end{equation}where $1 \leq j_1
< j_2 < \cdots < j_{2k} \leq n.$ By Eq.(22) we conclude from
Eq.(27) that $\lambda_{\epsilon_1 \cdots \epsilon_n} =0$ whenever
$\{\epsilon_{j_1}, \epsilon_{j_2},\ldots, \epsilon_{j_{2k}} \}$
contains odd number of $1.$ Therefore, besides $\lambda_{0 \cdots
0}$ and $\lambda_{1 \cdots 1},$ one has that $\lambda_{\epsilon_1
\cdots \epsilon_n} =0.$ This concludes Eq.(26) with $a= \lambda_{0
\cdots 0}$ and $b= \lambda_{1 \cdots 1}.$

(iii) Finally, we prove that if a state $| \varphi \rangle$ of $n$
qubits with $n \geq 3$ satisfies Eq.(7), then
\begin{equation}| \varphi
\rangle = \frac{1}{\sqrt{2}}\left (e^{i \phi} | 0\cdot \cdot \cdot
0 \rangle_n + e^{i \theta} i| 1\cdot \cdot \cdot 1 \rangle_n
\right ),\end{equation}for some $0 \leq \phi, \theta \leq 2 \pi.$

By Eq.(3), one concludes that Eq.(7) is equivalent to that ${\cal
M}_n |\varphi \rangle = 2^{n-1} |\varphi \rangle.$ In this case,
$|\varphi \rangle$ satisfies Eq.(25) and hence is of the form
Eq.(26). By using Eq.(2) we have that$$A'_1A_2\cdots A_n |\varphi
\rangle = |\varphi \rangle.$$Hence, by Eqs.(23) and (24) one has
that$$ia e^{-i(\alpha_1 + \alpha_2+\cdots +\alpha_n )} =b,$$that
is, $|a|=|b|=\frac{1}{\sqrt{2}}.$ Therefore, Eq.(28) holds true.

Now, denote by $V_j$ the unitary transform from the original
$\sigma^j_z$-representation to $A''_j$-representation on the $j$th
qubit, i.e., $V_j |0 \rangle = |0 \rangle_j$ and $V_j |1 \rangle =
|1 \rangle_j,$ and define$$U_1 = V_1 \left ( \begin{array}{ll} e^{
i\phi} & 0
\\~ 0 & 1 \end{array} \right ),U_2 = V_2 \left(
\begin{array}{ll}1 & ~0
\\ 0 & e^{i \theta} \end{array} \right ),$$and $ U_j =V_j$ for
$3 \leq j \leq n.$ Then $U_j$ are all unitary operators on ${\bf
C}^2$ so that Eq.(8) holds true and the proof is complete.

In conclusion, by using some subtle mathematical techniques we
have shown that the maximal violation of Mermin's inequality of
$n$ qubits only occurs for the states obtained from the GHZ states
by local unitary transformations. The key point of our argument
involved here is by using the certain algebraic properties that
Pauli's matrices satisfy, which is based on the determination of
local spin observables of the associated Bell operator. The method
involved here is simpler (and more powerful) than one used in [12]
and can be applied to the Bell-Klyshko inequality [9]. Note that
both Mermin's and Bell-Klyshko's inequalities are special
instances of all-multipartite Bell-correlation inequalities for
two dichotomic observables per site derived by Werner-Wolf and
\.{Z}ukowski-Brukner [15]. This seems to suggest that the
statement holds true for each of the Bell-correlation inequalities
for two dichotomic observables per site. It is argued that
maximally entangled states should maximally violate Bell-type
inequalities [8] and all states obtained by local unitary
transformations of a maximally entangled state are equally valid
entangled states [16], we conclude that the Bell-correlation
inequalities for two dichotomic observables per site characterize
the GHZ states as maximally entangled states of $n$ qubits.
Finally, we remark that the W state$$|W \rangle =
\frac{1}{\sqrt{n}}\left (|10\cdots 0 \rangle + |010\cdots 0
\rangle +\cdots + |0\cdots 01 \rangle \right )$$cannot be obtained
from the GHZ states by a local unitary transformation and hence
does not maximally violate the Bell-correlation inequalities for
two dichotomic observables per site whenever the above conjecture
holds true, although it is a ``maximally entangled" state in the
sense described in [17]. Hence, from the point of view on the
violation of EPR's local realism, the W state might not be
regarded as a ``maximally entangled" state and we need some new
ideas for clarity of the W state [18].

%\newpage %Just because of unusual number of tables stacked at end
% Produces the bibliography via BibTeX.


\begin{thebibliography}{**}
\bibitem{1}J.S.Bell, Physics (Long Island City, N.Y.){\bf 1},
195(1964).
\bibitem{2}A.Einstein, B.Podolsky, and N.Rosen, Phys. Rev. {\bf
47}, 777(1935).
\bibitem{3}N.Gisin, Phys.Lett. A {\bf 154}, 201(1991).
\bibitem{4}J.F.Clauser, M.A.Horne, A.Shimony, R.A.Holt,
Phys.Rev.Lett. {\bf 23}, 880(1969).
\bibitem{5}D.M.Greenberger, M.A.Horne, and A.Zeilinger, in {\it
Bell's Theorem, Quantum Theory, and Conceptions of the Universe,}
edited by M.Kafatos (Kluwer, Dordrecht, 1989), p.69; N.D.Mermin,
Phys. Today {\bf 43}, No. 6, 9(1990); N.D.Mermin, Am. J. Phys.
{\bf 58}, 731(1990).
\bibitem{6}N.Gisin, G.Ribordy, W.Tittle, and H.Zbinden,
Rev. Mod. Phys. {\bf 74}, 145(2002); M.Nielsen and I.Chuang, {\it
Quantum Computation and Quantum Information}
(Cambridge University
Press, Cambridge, England, 2000).
\bibitem{7}A search for ``maximally entangled states" in tittle on
www.arxiv.org/quant-ph yields 36 results.
\bibitem{8}N.Gisin and H.Bechmann-Pasquinucci, Phys. Lett. A {\bf
246}, 1(1998).
\bibitem{9}Z.-Q.Chen, quant-ph/0402007.
\bibitem{10}References can be found in: A.Cabello, quant-ph/0012089.
\bibitem{11}N.D.Mermin, Phys.Rev.Lett., {\bf 65}, 1838(1990).
\bibitem{12}S.L.Braunstein, A.Mann, and M.Revzen, Phys.Rev.Lett.
{\bf 68}, 3259(1992); G.Kar, Phys.Lett. A {\bf 204}, 99(1995);
J.L.Cereceda, Phys. Lett. A {\bf 212}, 123(1996); J.L.Cereceda,
Phys. Lett. A {\bf 286}, 376(2001).
\bibitem{13}L.Hardy, Phys.Lett. A {\bf 160}, 1(1991);
D.N.Klyshko, Phys. Lett. A {\bf 172}, 399(1993); A.V.Belinskii,
D.N.Klyshko, Phys. Usp. {\bf 36}, 653(1993).
\bibitem{14}W.Pauli, Zeit.Physik, {\bf 43}, 601(1927).
\bibitem{15}R.F.Werner and M.M.Wolf, Phys.Rev. A {\bf 64},
032112(2001); M.\.{Z}ukowski and \v{C}.Brukner, Phys.Rev.Lett.,
{\bf 88}, 210401(2002).
\bibitem{16}N.Linden, S.Popescu, Fortsch.Phys. {\bf 46},
567(1998).
\bibitem{17}W.D\"{u}r, G.Vidal, and J.I.Cirac, Phys.Rev. A {\bf
62}, 062314(2000).
\bibitem{18}A.Sen(De), U.Sen, M.Wie\'{s}niak, D.Kaszlikowski, and
M.\.{Z}ukowski, Phys.Rev. A {\bf 68}, 062306(2003).
\end{thebibliography}
\end{document}